\documentclass{article}
\usepackage{psfrag,latexsym,amssymb,amsmath,natbib,color,epstopdf,lineno}
\usepackage{graphicx}
\def\bra{\langle}
\def\ket{\rangle}
\def\p{\partial}

\def\beq{\begin{equation}}
\def\eeq{\end{equation}}
\def\la{\label}
\def\ii{{\rm i}}

\def\figpath{./}
\def\spacce#1{\hskip #1pt}
\def\drawline#1#2{\raise 2.5pt\vbox{\hrule width #1pt height #2pt}}

\def\bdash{\hbox{\drawline{5.8}{.5}\spacce{2}}}

\def\trian{\raise 1.25pt\hbox{$\scriptstyle\triangle$}\nobreak}

\def\dtrian{\raise 1.25pt\hbox%
{$\scriptscriptstyle\bigtriangledown$}\nobreak}

\def\squar{\raise 1.25pt\hbox{$\scriptstyle\Box$}\nobreak}

\def\diamon{\raise 1.25pt\hbox{$\scriptstyle\diamond$}\nobreak}

\newcommand{\soliddtrian}{$\blacktriangledown$\nobreak}

\def\linedtri1{\hbox{\bdash\hspace{-1.6mm}$\bigtriangleup$\hspace{-0.8mm}\bdash}\nobreak}
\def\soliddtrian1{$\blacktriangledown$\nobreak}
\def\solidrtrian2{$\blacktriangleright$\nobreak}
\def\solidltrian3{$\blacktriangleleft$\nobreak}

\title{The streaks of wall-bounded turbulence need not be long}

\author{Javier Jim\'enez\\
School of Aeronautics, U. Polit\'ecnica Madrid, 28040 Madrid Spain}
\date{\today}

\begin{document}
\maketitle
\begin{abstract}
The effect of damping the longest streaks in wall-bounded turbulence is explored using
numerical experiments. It is found that long streaks are not required for the
self-sustenance of the bursting process, which is relatively little affected by their
absence. In particular, there are turbulence states in which the fluctuations of the
streamwise velocity have approximately the same length as the bursts, and are thus
presumably associated with the bursts themselves, while the burst structure is
essentially indistinguishable from flows in which longer velocity fluctuations are present.
This suggests that the long streaks found in unmodified flows may be byproducts,
rather than active parts of the energy generation cycle.
\end{abstract}


\section{Introduction}

\begin{figure}
\centering%
\includegraphics[width=.90\textwidth,clip]{\figpath 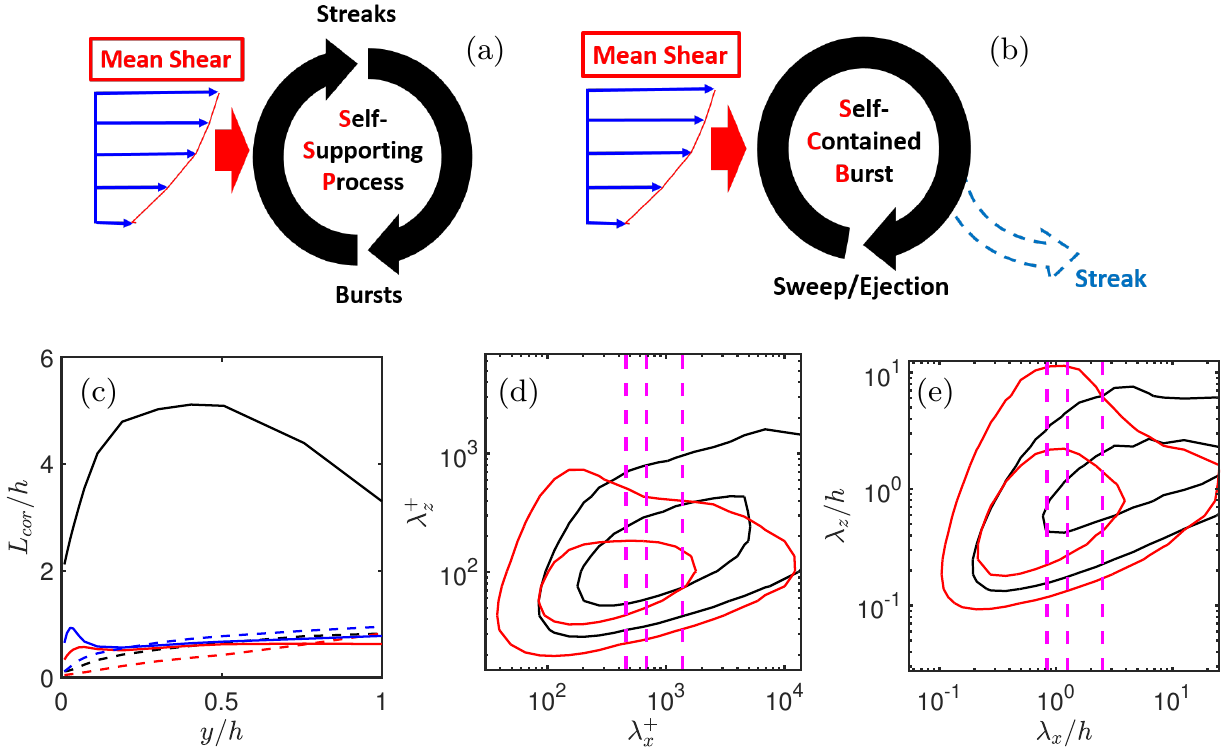}%
\caption{%
(a) The classical self-supporting cycle of wall turbulence.
(b) An alternative model in which the burst is self-contained and generates streaks
as by-products.
(c) Correlation lengths of the three velocity components, defined by the region of
the two-point velocity correlation function at a given height in which $C_{**}>0.1$.
Black, streamwise velocity; red, wall-normal;
blue, spanwise. Solid lines are streamwise correlations, and dashed lines are spanwise.
(d,e) Two-dimensional wall-parallel premultiplied spectra, $\Phi_{**}=k_x k_z E_{**}$.
Colours as in (c). Contours contain 50\% and 90\% of the spectral mass. The vertical
dashed lines are the limits of the damping in the simulations D10, D20 and D30 in table
\ref{tab:cases}, from right to left.
(d) Averaged over the buffer layer, $y^+\in (0, 60)$. 
(e) Averaged over the outer layer, $y/h\in (0.4, 0.7)$. 
(c-e) Are from F00 in table \ref{tab:cases}.
}
\label{fig:550spec}
\end{figure}

Elongated streaks of the streamwise velocity have long been recognised as important features
of wall-bounded turbulence \citep{kli:rey:sch:run:67}, and it was soon proposed that their
instability is the cause of the intermittent `bursting' that was later shown to be a dominant
contributor to the cross-stream transfer of momentum and to the generation of turbulent drag
\citep{lu:wil:73}. While this causal relation was initially hypothetical, detailed mechanisms
were eventually suggested \citep{swe:black:87}, and more concrete evidence appeared. For
example, \cite{jim:moi:91} showed that a minimal self-sustaining unit of plane Poiseuille
turbulence consists of a streak that is infinitely long when compared to the
minimal computational box, and a pair of shorter quasi-streamwise
vortices, whose intensity evolves in counter-phase with the streak. The kinematics of this
unit was made clearer in plane Couette flow by \cite{Hamilton95}, and eventually became
codified in a turbulence regeneration cycle in which the vortices generate the streaks by
deforming the mean velocity profile, and the streaks create the vortices by some sort of
instability \citep{jim94,Hamilton95}. \cite{flo:jim:10} extended minimal simulations to the
logarithmic layer and, by analogy with the viscous-layer results, concluded that a similar
regeneration cycle is active across the shear-dominated part of the channel, with the
streamwise vortices substituted by more generic transient bursts mainly involving
the cross-stream velocities, as in figure \ref{fig:550spec}(a). Although seldom made
explicit, it was usually implied that the arrows in this figure represent time. Reduced
dynamical models by \cite{Waleffe97} and others also typically include infinitely long
streaks and shorter vortices, and \cite{jim:pin:99}, by modifying different terms of the
evolution equations in turbulent channels at low Reynolds numbers, showed that both
components are required. Damping the streaks leads to the decay of the vortices, and vice
versa, and the result in both cases is laminarisation of the flow. Exact travelling waves
and other invariant solutions that mimic the near-wall structures also include an infinitely
long streak of the streamwise velocity and quasi-streamwise vortices
\citep{Nagata90,KawEtal12}.

Although no linear instability of the mean velocity profile is known for wall-bounded
turbulence \citep{reytied67}, each of the branches of the regeneration cycle in figure
\ref{fig:550spec}(a) can be modelled by transient, non-modal, linear amplification
processes. The bursts generate the streaks by linear advection of the mean profile, and the
bursts are amplified by the shear, through a tilting process in which eddies at different
distance from the wall are reinforced when they overtake each other \citep{orr07a,jim:13a}.
The optimum transient growth of linearised perturbations of the mean velocity profile
results in streaks and bursts compatible with those observed in experiments
\citep{but:far:93,ala:jim:06,jim:13a}, and \cite{vaug:etal:15}, among others, have presented
plausible self-sustaining models of wall turbulence based on time-dependent infinitely long
streaks. All these mechanisms are nonlinear in the sense of requiring initial
perturbations of sufficiently high amplitude to self sustain. Bursts are essentially linear
and grow by drawing energy from the mean shear \citep{jim:13a}, but infinitesimally small
initial perturbations only create infinitesimally weak bursts, which then result in
infinitesimally weak streaks. Only sufficiently strong streaks are unstable
\citep{Schoppa02}, and no linear self-sustaining mechanism is known.

Bursts and streaks have been extensively studied in turbulence simulations, both in the
viscous and in the logarithmic layer. Down- and up-drafts (the sweeps and ejections
that form the bursts) are statistically found in pairs consistent with the short
quasi-streamwise rollers that substitute the above-mentioned vortices outside the buffer
layer. They sit at the interface between longer high- and low-velocity streaks
\citep{loz:flo:jim:12}. Their lifetime is controlled by the shear of the mean velocity
profile \citep{loz:jim:14} and, when their evolution is conditionally averaged with respect
to the maximum intensity of the burst, the progressive tilting of the Orr mechanism is
clearly seen \citep{encinar:20}. \cite{jim18} reviews these observations and their
relation with theory.

Considering the above evidence, it may be surprising that the rest of this paper is
dedicated to showing that long streaks are not essential features of wall-bounded
turbulence, which can be maintained by short fluctuations of the streamwise velocity
presumably associated with the bursts themselves, and that the relation between bursts and
streaks is closer to the one in figure \ref{fig:550spec}(b), with a self-contained burst
that only generates the observed longer streaks as by-products. Although it may be debatable
whether flows without long streaks can be described as `wall turbulence', as is otherwise
the case for most of the artificial systems used to analyse turbulence, we will show that
the structure of the transverse velocity components of the remaining short scales differs
little from unmodified flows. There are three related questions in this respect. The first
one, whether infinitely long streaks {\em can} become unstable at streamwise wavelengths
comparable to the observed bursts, has been amply answered in the affirmative by the authors
mentioned above, and by many others. The second question is whether this `direct cascade',
from long streaks to shorter bursts, is a {\em necessary} part of the turbulent cycle, and
the third one is whether, even if this downward cascade is shown not to be a required part
of the cycle, it still is an {\em important} ingredient of turbulence dynamics. We argue
below that the answer to the second question is negative, and that the third question is
probably only true to the extent that similarly sized fluctuations of the three velocity
components must be present in self-sustained bursts, because bursts exclusively involving
velocities in the cross plane are equivalent to unforced two-dimensional turbulence and must
decay. A fourth question, related to the `inverse cascade' by which short bursts may create
long streaks, will be left unanswered.

In fact, the evidence for a close connection between the bursts and the much longer streaks
is not strong. Figure \ref{fig:550spec}(c) shows that their dimensions are very different.
The same-height correlation lengths in the figure underestimate the actual dimensions, which
increase when correlations at different heights are taken into account
\citep{sil:jim:mos:14}, but the ratio among structures is not modified. If we denote the
streamwise, wall-normal, and spanwise directions by $x$, $y$ and $z$, respectively, and the
corresponding velocity components by $u$, $v$ and $w$, the three-dimensional correlations of
the transverse velocities, $v$ and $w$, are approximately isotropic in the cross plane,
$\Delta_z/y\approx 1.5$, and only moderately elongated streamwise, $\Delta_x/y\approx 3$.
The correlation of the streamwise velocity is much more elongated, with spanwise dimensions
similar to those of the transverse velocities, but with $\Delta_x/y\approx 15$--30. The same
difference emerges from spectra (see figure \ref{fig:550spec}d,e) and from the structural
analysis of the bursts by \cite{loz:flo:jim:12}. On the other hand, when the thresholding
method used for bursts is applied to the fluctuations of $u$, their dimensions are only
$\Delta_x/y \approx 10$ \citep[see figure 22 in][]{jim18}, suggesting that the correlations
of $u$ represent a higher level of organisation of shorter units, each of them presumably
associated with a burst or with a group of bursts. Moreover, the $u$-correlations of
boundary layers are two or three times shorter than channels, but the two flows have very
similar statistics below the outer intermittent region \citep{sil:jim:mos:14}, and, if we
denote by $h$ the channel half-width or other suitable flow thickness, the longest $u$
correlations in channels have $\Delta_x/h\approx 7$--15, while \cite{mos:kim:man:99} and
\cite{lozano14} showed that shorter computational boxes, with lengths $L_x=2\pi$, give good
statistics.

However, all these simulations, which are typically spectral, contain the $k_x=0$
longitudinal wavenumber that corresponds to an infinitely long streamwise wavelength, and it
could be argued that these infinitely long features are responsible for maintaining the
flow, as in some of the theoretical models mentioned above. This was explicitly
tested by \cite{jim:pin:99}, who forcibly zeroed all the Fourier components with $k_x=0$ of the
wall-normal vorticity, preventing the formation of infinitely long spanwise
inhomogeneities of $u$. The flow laminarised in a computational box with $L_x^+ =500$,
where the `+' superscript denotes wall units defined from the friction velocity, $u_\tau$,
and from the kinematic viscosity $\nu$. Repeating the experiment in a longer box
$(L_x^+=1200)$ failed to laminarise the flow, which could only be done by zeroing the first
two wavenumbers ($k_x=0$ and $k_x=2\pi/L_x$), thus only retaining uniform streaks with
$\lambda_x^+\le L_x^+/2=600$. These experiments were performed at low Reynolds numbers,
$h^+\approx 180$, and were interpreted to mean that features of at least this length are
required for the survival of the near-wall buffer region.

The present paper revisits the experiments of \cite{jim:pin:99}, which can now be performed
at higher Reynolds numbers and in larger boxes, and analysed in the context of the 
structural information gained since then. The numerical details are given in
\S\ref{sec:simul}, results are presented in \S\ref{sec:results}, and are discussed and
summarised in \S\ref{sec:conc}.

  
\begin{table}
  \begin{center}
    \def~{\hphantom{0}}
    \begin{tabular}{lccccccll}
      Case  &  $Re_\tau$ & $\Delta x^+$ &$\Delta z^+$ & $\Delta y^+_{max}$ & 
         $\lambda_{xf}/h$ & $\lambda_{xf}^+$ & Damping height & Result\\[3pt]
      F00                & 550 & 8.9 & 5.5 & 6.7  & $\infty$  & $\infty$  & 
                                                            \multicolumn{2}{c}{\cite{juanc03}}\\
      D10                & 452 & 7.4 & 3.7 & 5.5  & 2.28 & 1033 & Full channel & turbulent   \\
      D15                & 381 & 6.2 & 3.1 & 4,7  & 1.57 & 598 & Full channel & turbulent   \\
      D20                & 345 & 5.6 & 2.8 & 4.2  & 1.20  & 413 & Full channel & transitional\\     
      D30                & 252 & 5.6 & 2.8 & 4.2  & 0.81  & 280 & Full channel & laminar\\
     D10out            & 453 & 7.4 & 3.7 & 5.6  & 2.28  & 1035 & Above $0.18h$ & turbulent\\
     D10in              & 472 & 7.7 & 3.8 & 5.8  & 2.28  & 1078 & Below $0.18h$ & turbulent\\
  \end{tabular}
\caption{Parameters of the large-box simulations. The size of the doubly periodic
computational box is $L_x\times L_z=(8\pi\times 4\pi)h$, and $U_bh/\nu=10^4$, where $U_b$ is
the bulk velocity. All simulations use the same collocation grid ($1536\times 257\times
1536$ in $x,y,z$). The friction Reynolds number is measured at the end of each simulation,
and $\lambda_{xf}$ is the longest undamped wavelength. Wall units for D30 are based on D20.
 }%

    \label{tab:cases}
  \end{center}
\end{table}

\section{Numerical experiments}\la{sec:simul}

We analyse simulations of pressure-driven incompressible turbulent channels between flat
plates separated by a distance $2h$, in doubly periodic computational boxes whose streamwise
and spanwise periods are $L_x$ and $L_z$, respectively. The code is Fourier-spectral in
$(x,z)$ and Tchebychev-spectral in $y$, and follows closely \cite{KMM87}. More details are
found in \cite{juanc03}, whose simulation at $Re_\tau\equiv h^+\approx 550$ is used here as
a reference. The code integrates evolution equations for $\nabla^2 v$ and for the
wall-normal vorticity, $\omega_y$, from where other variables are obtained using continuity.
The volume flux per unit span, $2hU_b$ is kept constant. To control the spanwise
inhomogeneity of the streamwise velocity, the variable to be modified is $\omega_y$, which is
approximately equal to $\p_z u$ for long and narrow features. Following \cite{jim:pin:99},
this modification is implemented by explicitly zeroing at each time step all the long harmonics of the
Fourier expansion $\omega_y (x,y,z) =\sum\sum\widehat\omega_y (k_x,y,k_z) \exp[\ii (k_x x
+k_z z)]$, so that
\beq
\widehat\omega_y (k_x,y,k_z) \to 0,\quad\mbox{for all $k_z$, and}\; k_x<  2\pi/\lambda_{xf}.
\la{eq:damp1}
\eeq
This includes the infinitely long modes with $k_x=0$. A few simulations were run in
which the damping is only applied above or below a given distance from the wall, $y_f$,
\beq
\widehat\omega_y (k_x,y,k_z) \to \widehat\omega_y (k_x,y,k_z) F(y),
\quad\mbox{for all $k_z$, and}\; k_x< 2\pi/\lambda_{xf},
\la{eq:damp2}
\eeq
with
\beq
F(y)=\tfrac{1}{2} \left[1\pm 
      \tanh\frac{y-y_f}{\sigma_f}\tanh\frac{2-y-y_f}{\sigma_f} 
\right], \qquad y\in (0,2).
\la{eq:damp3}
\eeq
The plus sign damps the vorticity below $y_f$, and the minus sign damps it above that level.
In our simulations, $y_f=0.2$ and $\sigma_f=0.05$. No modification is applied to the
$\nabla^2v$ equation. Other numerical parameters are summarised in table \ref{tab:cases},
and some wavelength limits are overlaid on the spectra in figure \ref{fig:550spec}(d,e).
It should be noted that, because of the choice of simulation variables, arbitrarily
manipulating them respects continuity.

The damping of the vorticity is equivalent to a body force (or torque) acting in planes parallel to
the wall, but because of homogeneity, and because it acts on velocity derivatives rather
than on the velocities themselves, its average cancels over time. The total stress profile
of all the simulations discussed below is linear.

\section{Results}\la{sec:results}
\begin{figure}
\centering
%
%
\includegraphics[width=.90\textwidth,clip]{\figpath 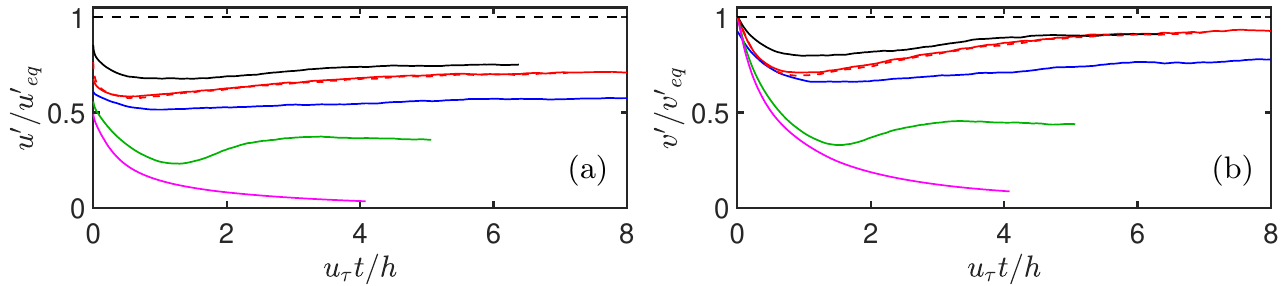}%
\caption{%
Temporal evolution of the box-averaged velocity fluctuation intensity in the full-box
simulations in table \ref{tab:cases}, normalised with the equilibrium values in
\cite{juanc03}. (a) Streamwise velocity. (b) Wall-normal velocity.
Solid red line, D10; solid blue, D15; solid green, D20; solid magenta, D30; dashed red,
D10out; solid black, D10in.
}
\label{fig:550cf}
\end{figure}

\begin{figure}
\centering
\includegraphics[width=.98\textwidth,clip]{\figpath 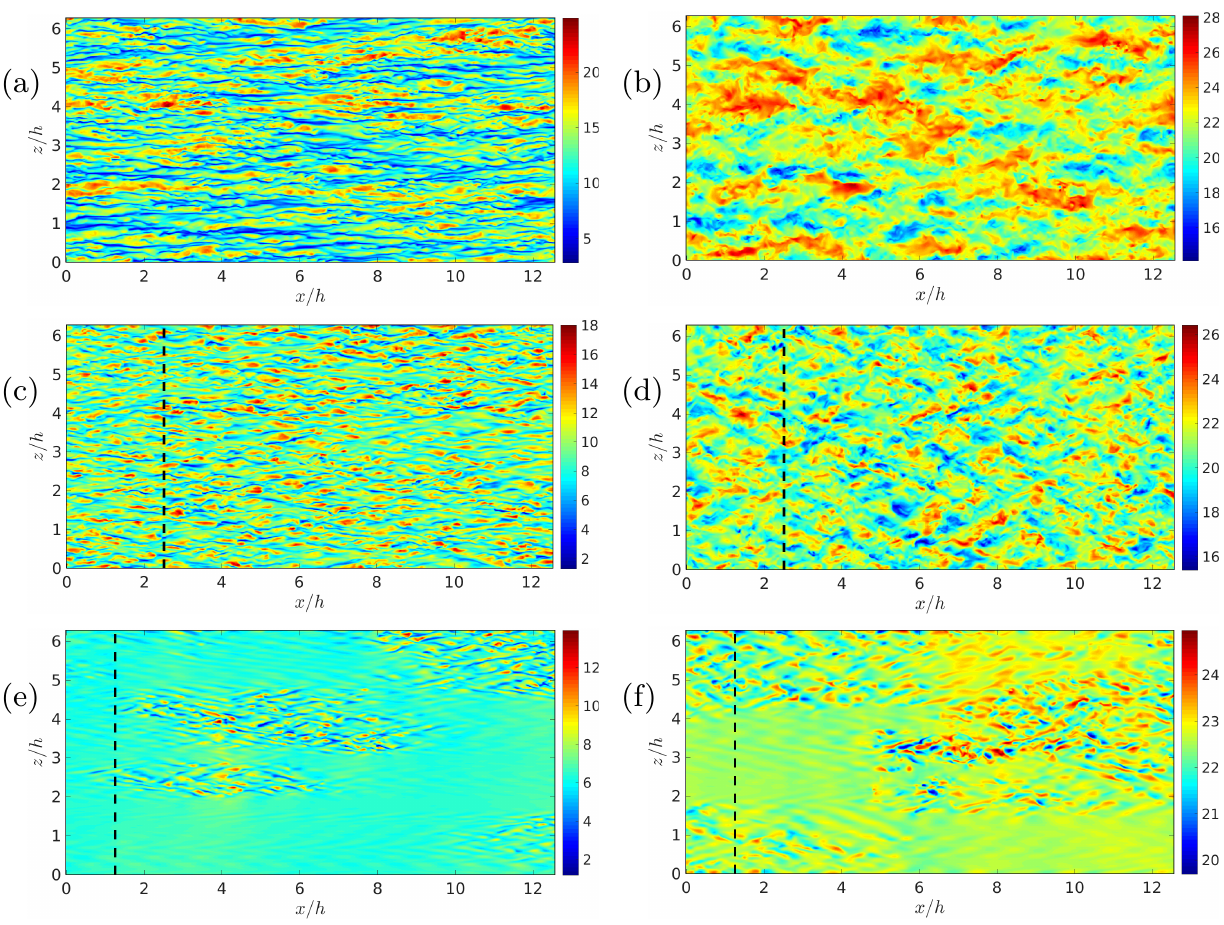}%
\caption{%
Wall-parallel sections of the streamwise velocity. The flow is from left to right, and all
normalisations use the friction velocity from \cite{juanc03}. Each panel contains one fourth
of the computational plane. (a,b) Undamped, F00. (c,d) D10. (e,f) D20. The distance from the
origin to the vertical dashed line is $\lambda_{xf}$. (a,c,e) $y^+=15$. (b,d,f) $y/h=0.346$.
}
\label{fig:550field}
\end{figure}

The temporal evolution of the $L_2$ norm of two velocity components of the damped
simulations is given in figure \ref{fig:550cf}, normalised with the equilibrium values of an
undamped reference simulation at the same nominal Reynolds number \citep[F00]{juanc03}. The
damped cases initially decay, but later recover and stabilise at a lower friction Reynolds
number than the reference. Case D30 laminarises completely, and does not recover. Figure
\ref{fig:550field} shows wall-parallel sections of the streamwise velocity in the buffer
$(y^+=15)$ and outer $(y/h\approx 0.35)$ regions of the reference case and of two surviving
damped cases, D10 and D20.

Case D20 contains a mixture of laminar and turbulent patches, reminiscent of transition, which
persists for as long as the case was run. Case D10, which filters only wavelengths longer
than $\lambda_x\approx 2.5h$, stays turbulent, but with a very different organisation from
the nominal case, whose streaky structure is substituted by a rhomboidal pattern. This
organisation is easy to rationalise, because the damping ensures that the streamwise average
of the velocity fluctuations vanishes over streamwise distances longer than $\lambda_{xf}$,
so that a positive fluctuation has to be followed by a negative one. But the most
interesting aspect of figure \ref{fig:550field}(c,d) is that the spanwise wavelength of the
velocity is approximately the same as in the nominal case, so that the angle of the
rhombuses increases as this wavelength increases with the distance from the wall. It is
intriguing that a weaker, but similar, rhomboidal pattern with an angle close to 45 degrees
is observed in the spanwise velocity of natural flows \citep{sil:jim:mos:14}. If the above
explanation applied to it, it would imply that the mean value of $w$ tends to cancel over
streamwise distances of the order of its spanwise scale. This is consistent with the lack of
long tails in its spectrum \citep{jim18}, and may be linked to the streak meandering
discussed by \cite{hutmar07}. It also has the right dimensions to be related to the bursts
described in the introduction.

\begin{figure}
\centering
\includegraphics[width=.90\textwidth,clip]{\figpath 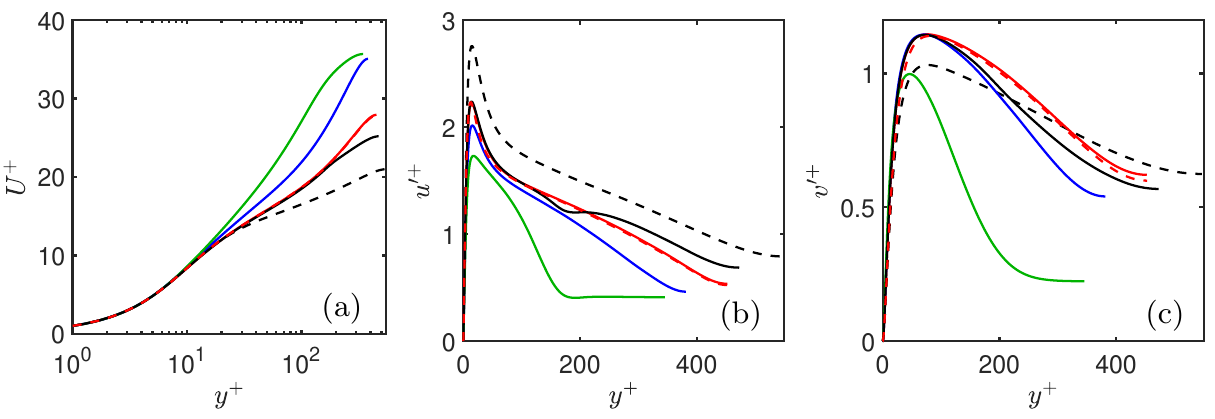}%
%
%
\caption{%
Mean and fluctuation velocity profiles of the filtered simulation in table \ref{tab:cases}. Colours as
in figure \ref{fig:550cf}. Dashed black line is F00. (a) Mean streamwise velocity, $U^+$. (b) $u'^+$. 
(c) $v'^+$.
}
\label{fig:550prof}
\end{figure}

\begin{figure}
\centering
\includegraphics[width=.90\textwidth,clip]{\figpath 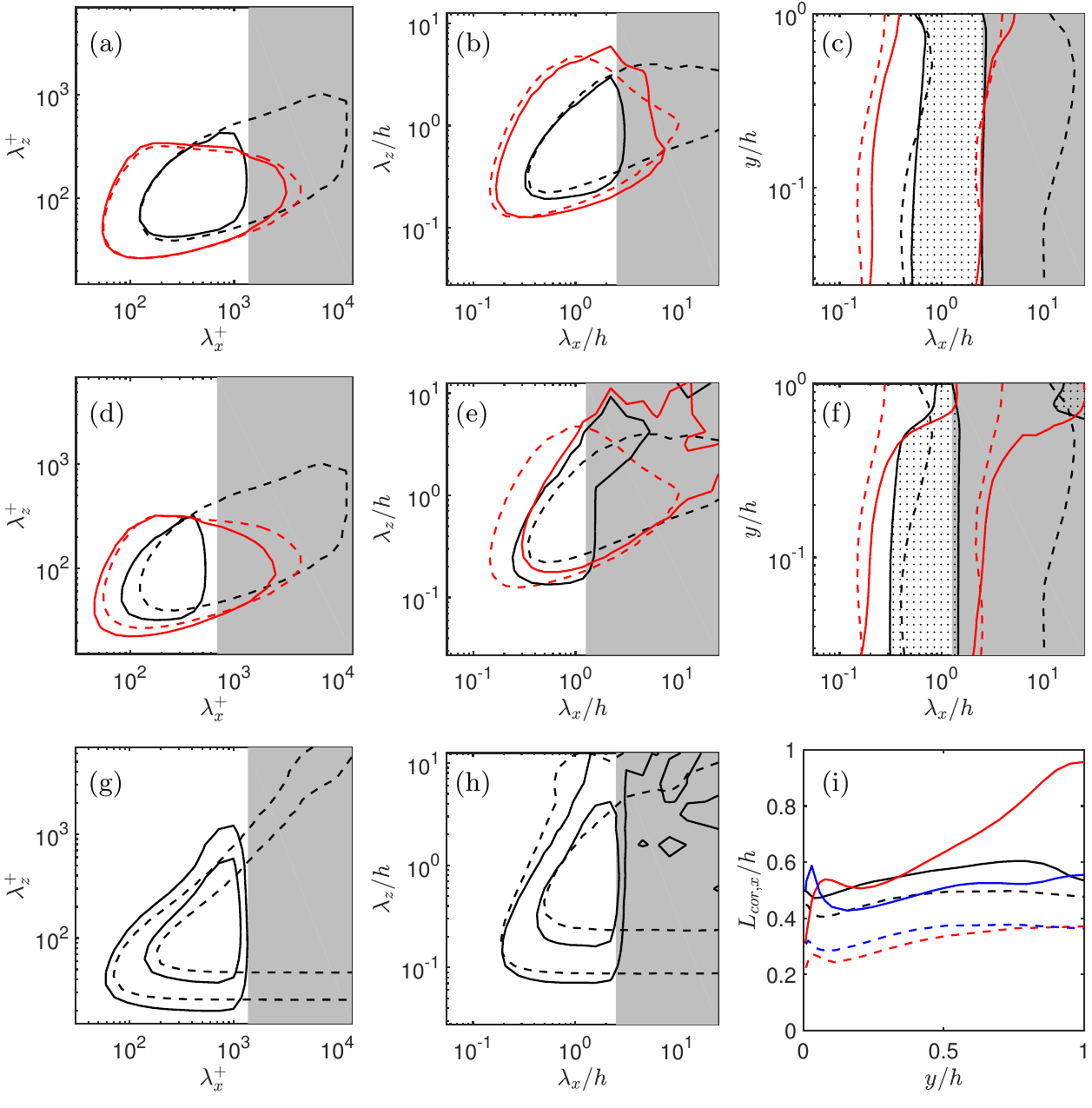}%
%
%
\caption{%
Premultiplied energy spectrum. The dashed contours are F00. The solid ones are
damped cases.  The damped region is in grey, and spectral contours contain 75\% of the spectral
mass. Black lines, $\Phi_{uu}$; red ones, $\Phi_{vv}$.
(a-c, g-i) D10. (d-f) D20. 
(a,b,d,e) $k_xk_z E$.
(g,h) $-E_{uv}/(E_{uu} E_{vv})^{1/2}$. Contours 0.1 and 0.4. Solid lines are D10;
dashed are F00.
(a,d,g) Averaged over the buffer layer, $y^+\le 60$, as in figure \ref{fig:550spec}(b). 
(b,e,h) Averaged over the outer  layer, $y/h\in(0.4,0.7)$, as in figure \ref{fig:550spec}(c).
(c,f) Isocontours of $k_x E_{\bullet\bullet}(k_x,y)/\bullet'^2(y)$, versus wavelength and
distance from the wall (one of the contours is filled for clarity).
(i) Streamwise correlation lengths of the three velocity components, defined by the
region of the two-point velocity correlation function at a given height in which
$C_{**}>0.1$. Black, streamwise velocity; red, wall-normal; blue, spanwise. Solid lines are
D10, and dashed ones are F00 that has been filtered to the same wavenumber as D10.
}
\label{fig:filtspec}
\end{figure}

The mean and fluctuation velocity profiles of the damped simulations are shown in figure
\ref{fig:550prof}, compared with the reference case. In all cases except the transitional
D20, the streamwise fluctuations are weaker than the reference case, while the wall-normal
ones are stronger. Although not shown, the behaviour of $w$ follows the same trend
as $v$. This is characteristic of controlled flows, because the relative magnitudes of 
$u$ and $v$ have to adjust themselves to generate the tangential Reynolds
stress used to define wall units. It is clear from the mean velocity profiles in figure
\ref{fig:550prof}(a) that D20 is basically laminar above $y^+\approx 200$, with no trace of
a logarithmic layer, but the behaviour of D10 is interesting. When the whole channel is
damped, as in D10, the mean velocity can be described as a logarithm with a low von
K\'arm\'an constant, reflecting the missing Reynolds stresses. The same is true when only
the outer layer is filtered, as in D10out, whose red dashed line is difficult to distinguish
from the solid line of D10. However, when only the inner layer is filtered, as in the black
solid line of D10in, the mean velocity and the fluctuation profile of $u'$ tend to recover
their turbulent behaviour in the undamped layer above $y/h=0.3\, (y^+\approx 150)$. This
supports the idea that the large scales of the outer part of the channel are not driven by
the wall.

The premultiplied spectra in figure \ref{fig:550spec}(b,c) suggest that the behaviour of the
flow depends on how much of the $v$ spectrum is damped. Case D10 barely modifies
$\Phi_{vv}$, either in the buffer or in the outer layer, but D20 damps a large fraction of
$v$, especially in the outer layer where figure \ref{fig:550prof} shows that turbulence has
died. The streamwise velocity spectrum, $\Phi_{uu}$, is heavily damped in both cases, but it
does not appear to have a strong effect on $v$ (or in $w$). This is confirmed by figure
\ref{fig:filtspec}, which compares spectra of damped and undamped simulations. In figure
\ref{fig:filtspec}(a-c), $\Phi_{uu}$ is heavily truncated in D10, but the remaining part
agrees well with a truncated version of the undamped spectrum. Up to a point, the same is
true of $\Phi_{vv}$, most likely because its truncation is minor. The truncation of
$\Phi_{vv}$ in the near-wall region of D20 is also moderate, but all the spectra are heavily
distorted in the outer layer of D20 in figure \ref{fig:filtspec}(e), where half of
$\Phi_{vv}$ is damped. In fact, although the contours in these figures are chosen to
represent a fixed fraction of the fluctuation energy (75\%), figure \ref{fig:550prof} shows
that the outer spectra in figure \ref{fig:filtspec}(e) correspond to weak residual fluctuations,
which are not turbulent. Another interpretation of these spectra, especially clear from the
$(\lambda_x, y)$ projections in figure \ref{fig:filtspec}(c,f), is that turbulence requires
that the fluctuations of $u$ be longer than those of $v$, but only by a small factor of two
or three.

While these spectra suggest that filtering the long wavelengths of $\omega_y$ has relatively
little effect on $v$, it is interesting to compare the shorter surviving scales with those
at similar wavelengths in a full channel. This is difficult in absolute terms, because it is
unclear which $u_\tau$ to use for individual wavenumbers, but a simple dimensionless
indicator of flow `sanity' is the structure parameter formed by the $u$--$v$ co-spectrum
normalised with the spectra of $u$ and $v$. This is shown in figure \ref{fig:filtspec}(g,h)
for D10. The long wavelengths that have been damped lose all correlations, and do not
contribute to the tangential Reynolds stress, but the structure parameter of the undamped
wavelengths is essentially unchanged with respect to the full channel, even near the edge of
the damped region. Figure \ref{fig:filtspec}(i) shows streamwise correlation lengths for the
damped channel and for a full one in which the damped wavenumbers have been zeroed during
postprocessing. It should be compared with the unfiltered results in figure
\ref{fig:550spec}(c). The size of the transverse velocities has changed little, but the long
streamwise correlations of $u$ have disappeared. There is very little effect on the spanwise
correlation lengths (not shown). The implication is that, while damping the transverse
velocities (and therefore the bursts) kills turbulence at the damped wavenumbers, the large
difference between the length of $u$ and $v$ (and $w$) in nominal turbulence is not required
for its survival.

\begin{table}
  \begin{center}
    \def~{\hphantom{0}}
    \begin{tabular}{lccccccccl}
      Case  &  $Re_\tau$ & $\Delta x^+$ &$\Delta z^+$ & $\Delta y^+_{max}$ & 
         $L_x^+$ & $L_z^+$ & $\lambda_{xf}/h$ & 
                $\lambda_{xf}^+$ & Result\\[3pt]
      F950               & 949 & 7.8 & 3.9 & 7.8  & 1490 & 745 & $\infty$  & $\infty$  & \cite{jim:13a}\\
      D950-0            & 830 & 6.8 & 3.4 & 6.8  & 1300&  651 & 1.57 & 1300 &  Turbulent  \\
      D950-1            & 555 & 4.5 & 2.3 & 4.5  & 870 &  435 &0.78  & 436 &  Turbulent \\     
      D950-2            & 241 & 4.5 & 2.3 & 4.5  & 870 & 435 & 0.52  & 291 & Laminar\\
  \end{tabular}
\caption{Parameters of the small-box simulations. The size of the doubly periodic computational box is
$L_x\times L_z=(\pi h/2\times \pi h/4)$, and $U_bh/\nu=19340$, where $U_b$ is the bulk
velocity. All simulations use the same collocation grid ($192\times 385\times 192$ in
$x,y,z$). The friction Reynolds number is averaged over the last half of each simulation, and
$\lambda_{xf}$ is the longest undamped wavelength. Wall units for D950-2 are based on D950-1. 
 }
   \label{tab:cases950}
  \end{center}
\end{table}

\begin{figure}
\centering
\includegraphics[width=.90\textwidth,clip]{\figpath 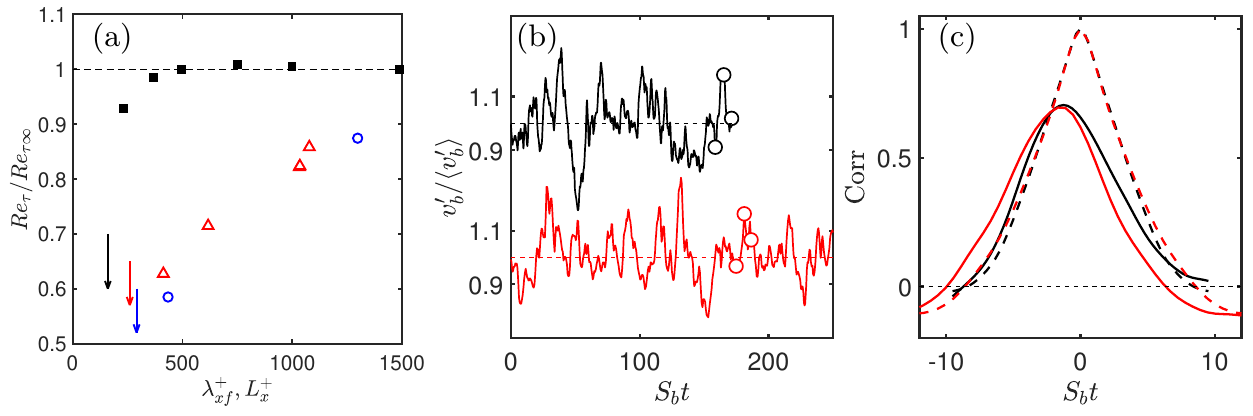}%
%
%
%
\caption{%
(a) Decay of the friction Reynolds number with the streamwise wavelength. Red triangles and
blue circles are cases in tables \ref{tab:cases} and \ref{tab:cases950}, respectively, plotted
versus $\lambda_{xf}^+$; solid squares are undamped channels with the same parameters as
F950 in table \ref{tab:cases950}, but with shorter boxes, plotted against $L_x^+$. The arrows mark
cases that laminarise, scaled in the wall units of the closest surviving simulation.
(b) Temporal evolution of the band-averaged $v$ fluctuation intensity, normalised with its
temporal mean, and offset for clarity. Top line is F950, and bottom one is D950-0. Symbols
mark the snapshots in figure \ref{fig:streaks}.
(c) Dashed lines are $C(v'^2_b, v'^2_b)$, and solid ones at $C(v'^2_b, u'^2_b)$. Black is
F950; red is D950-0. Time in (b,c) is normalised with the mean shear in the averaging band.
}
\label{fig:Redecay}
\end{figure}

To explore the dependence on the Reynolds number, and since the previous results suggest
that features longer than $\lambda_x \approx h$ do not have a strong effect on the dynamics
of the structures in the core of the spectrum, two sets of simulations were run in the
relatively small boxes used by \cite{jim:13a} to study the logarithmic layer (case F950 in
table \ref{tab:cases950}). According to \cite{flo:jim:10}, turbulence in these boxes should
be healthy below $y\approx 0.3 h,\, L_z \approx 0.24 h$. The first set of simulations,
detailed in table \ref{tab:cases950}, are damped ones in which the first few wavenumbers are
zeroed. The second set are undamped versions of F950 in which the streamwise length of the
box is progressively shrunk until turbulence is no longer sustained. This second set of
simulations retains the $k_x=0$ mode, and are represented by solid symbols in figure
\ref{fig:Redecay}(a).
  
The open symbols in this figure summarise the behaviour of the damped boxes, compared with
the corresponding undamped simulations. As the damping becomes more severe and the Reynolds
stress from the damped scales disappears, the friction Reynolds number decreases with
respect to its undamped value, and the flow laminarises for $\lambda_{xf}^+ \lesssim 400$.
We have already mentioned that \cite{jim:pin:99} were able to sustain turbulence for
$\lambda_{xf}^+\gtrsim 600$, but not for shorter wavelengths. This is also the limit found
by \cite{jim:moi:91} for the $L_x^+$ of minimal channels at $L_z^+ \approx 100,\,
Re_{\tau}\approx 180$, suggesting that the flow only fully laminarises when the box is too
short even for the smallest structures in the buffer layer. This is confirmed by the
collapse of figure \ref{fig:Redecay}(a) in wall units, although the limit at the
higher Reynolds number of the present simulations in closer to $\lambda_{xf}^+\approx 450$.
On the other hand, the behaviour of damped and undamped boxes is different. The undamped
simulations represented by the solid symbols in figure \ref{fig:Redecay}(a) retain their
infinitely long structures and the associated Reynolds stress, and the figure shows very
little degradation of $Re_\tau$ until they laminarise for very short $L_x$. The limit of
short and wide channels has been extensively explored by \cite{TohItan05} and coworkers, who
also find little degradation for boxes longer than $L_x^+ \approx 260$ at $Re_{\tau}= 137$.
Figure \ref{fig:Redecay}(a) confirms that a similar limit holds at higher Reynolds number.
It may be significant that the undamped limit is shorter than the damped one by a factor of
about two, which would confirm the similar suggestion from the spectra in figure
\ref{fig:filtspec}.

An advantage of small boxes is that the uppermost  band of wall distances for which
turbulence remains healthy is essentially minimal for the energy-containing eddies, and
their temporal evolution can be studied by tracking band-averaged integral quantities. For
the rest of this section we will use fluctuation intensities averaged over $y/h\in
(0.15-0.3)$, denoted by a `$b$' subscript.

Figure \ref{fig:Redecay}(b) displays the temporal behaviour of $v'_b$ in cases F950 and
D950-0 of table \ref{tab:cases950}, which only differ by the zeroing in the latter of
$k_x=0$. Both quantities burst intermittently with an approximately similar period, and with
an amplitude that is only slightly larger in the undamped case. The temporal evolution can
be quantified by the temporal correlation, which is defined for any two variables $a$ and
$b$ as,
\beq
C(a,b; t) = \frac{\bra a(s) b(s+t)\ket_s}{(\bra a^2\ket_s \bra b^2\ket_s)^{1/2}}, 
\la{eq:corrt}
\eeq
where the average $\bra\ket_s$ is taken over time and over the two sides of the channel. 

\begin{figure}
%
\centering
\includegraphics[width=.98\textwidth,clip]{\figpath 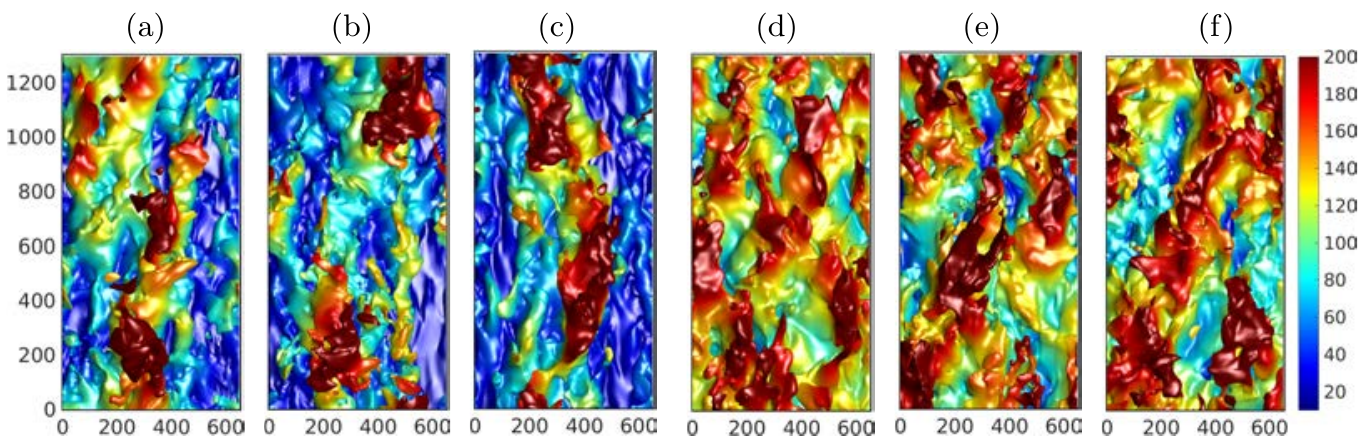}%
\caption{%
Wall-parallel views of small-box simulations. Colours are distance from the wall of the $U^+=18$ 
isosurface. Flow is from the bottom upwards, and lengths are in wall units.  
(a-c) F950. Time between consecutive frames , $u_\tau \Delta t/h\approx 0.52$. 
(d-f) D950-0. Time between consecutive frames,  $u_\tau \Delta t/h\approx 0.30$. 
}
\label{fig:streaks}
\end{figure}

Figure \ref{fig:Redecay}(c) shows correlations of $u'^2_b$ and $v'^2_b$, and should be
compared with figure 6 in \cite{jim:13a}. The correlation $C(v'^2_b, v'^2_b)$ is symmetric
and agrees very well for the two flows, again suggesting that the dynamics of $v$
is unaffected by damping longer $u$ structures. \cite{jim:13a} shows that this evolution
represents well an \cite{orr07a} burst. As mentioned in the introduction, such bursts also
typically involve $u'_b$, but the correlation $C(v'^2_b, u'^2_b)$ is not necessarily
symmetric, and depends on the location of the burst with respect to pre-existing
$u$ fluctuations. In agreement with \cite{jim:13a}, $C(v'^2_b,u'^2_b)$ is tilted
towards negative times in F950, suggesting that the streak weakens as the burst develops
\citep[curiously, in apparent contradiction with the idea that bursts generate the streaks,
but see figure 7c in][]{encinar:20}. This is not substantially changed for the
damped flow.

Snapshots of flow fields in small boxes are presented in figure \ref{fig:streaks}. The three
leftmost panels are the undamped case F950, and the three rightmost ones are D950-0. The
figures map the distance from the wall of the $U^+=18$ isosurface, which has been chosen
because its average position is $y^+\approx 100$ in both flows. In practice, it ranges over
$y^+\in(15-200)$, and the colour code is uniform among panels. The three snapshots in each
flow are chosen to span a `burst' of $v'_b$, with its peak at the central panel. Red
areas are low-velocity regions, and it is interesting that the maps of D950-0 show the same
rhomboidal pattern as the large-box damped flows in figure \ref{fig:550field}(c-f). The burst in
figure \ref{fig:streaks}(e) forms when the symmetry of the rhombuses is broken and one
orientation dominates over the other. It is tempting to see a similar process in the undamped
flow fields, where the breakdown takes places as the tail of the streak `catches up' with
itself, although it should be remembered that these snapshots cover a spatially periodic
domain, and that their behaviour may partly be an artefact of periodicity.

\section{Discussion and conclusions}\la{sec:conc}

This note can be seen as a further step in the simplification of minimal models for wall
turbulence. We have shown that there are turbulence states in which bursting takes
place without long structures of the streamwise velocity, while the undamped
wavelengths burst in essentially the same way as in unmodified flows. This suggests that
the long correlations of the streamwise velocity found in the latter are not 
central to the dynamics, and that models are possible in which the reference to streaks in figure
\ref{fig:550spec}(a) is absent. Our results can even be understood to mean that streaks are
not required for the maintenance of wall-bounded turbulence, since there is little that can
be interpreted as a streak in figures \ref{fig:550field}(c-f) or \ref{fig:streaks}(d-f). In
fact, the length limits in figure \ref{fig:Redecay}(a) imply that the minimal dimensions of the
fluctuations of $u$ are $\lambda_x^+\times \lambda_z^+\approx 450 \times 100$,
which is much closer to a minimal burst than to an elongated streak. The spectra in figure
\ref{fig:filtspec} show that the damped wavenumbers do not contribute to the tangential
Reynolds stress, leading to a lower effective $Re_\tau$, but that the undamped modes of the
manipulated simulations are little affected by the absence of the longer wavelengths. This
is particularly clear in the agreement of the dimensionless spectral structure parameters in
figure \ref{fig:filtspec}(g,h).

This suggests that the classical sketch of the self-sustaining process in figure
\ref{fig:550spec}(a) may be substituted by the (solid) one-loop process in figure
\ref{fig:550spec}(b), at least in the manipulated flows. The dashed arrow in this figure
reflects that, in this model, the long streaks found in natural flow would be by-products of
the shorter bursts. This opens the question of how bursts are restarted, because, as was the
case of figure \ref{fig:550spec}(a), the new figure portraits a transient process. The hope
is that the new system might be easier to model than the previous one, because one of the
components of the latter is missing. A second question is what the role of the long streaks
in maintaining turbulence may be. It is unlikely that they have none, because they generate
Reynolds stresses at their wavelengths, and the decay of $Re_\tau$ in figure
\ref{fig:Redecay}(a) shows that their contribution to the total stress is substantial
\citep[see also][]{liu:adr:han:01}. What we have shown is that they are not essential
components of the bursting cycle, which continues relatively undisturbed when they are
removed. But they are present in unmodified flows, and may participate in some other
important process. The geometrical collocation of bursts with the interface between streaks
is well documented \citep{loz:flo:jim:12}, and the asymmetry of the $C(v'^2_b, u'^2_b)$
correlation in figure \ref{fig:Redecay}(c) suggests that, in unmodified flows, bursts appear
within streaks that later weaken. Lastly, the related `reverse-cascade' question of how
short transient bursts create long and long-lasting streaks, remains open.

\vspace{1ex}
This work was supported by the European Research Council under the Caust grant
ERC-AdG-101018287. 
%
\bibliographystyle{jfm}

\end{document}